\documentclass[amsmath,amssymb,twocolumn]{revtex4}
\usepackage[parfill]{parskip}    % Activate to begin paragraphs with an empty line rather than an indent
\usepackage{graphicx}
\usepackage{amssymb}
\usepackage{epstopdf}
\usepackage{color}

\usepackage{graphicx}
\usepackage{pdfpages}
\begin{document}

\title{Black hole shadow of a rotating polytropic black hole by the Newman--Janis algorithm
without complexification}

\author{Ernesto Contreras 
\footnote{econtreras@yachaytech.edu.ec}} 
\address{School of Physical Sciences \& Nanotechnology, Yachay Tech University, 100119 Urcuqu\'i, Ecuador\\}

\author{J. M Ramirez--Velasquez {\footnote{jmramirez@yachaytech.edu.ec}}}
\address{School of Physical Sciences \& Nanotechnology, Yachay Tech University, 100119 Urcuqu\'i, Ecuador\\}

\author{\'Angel Rinc\'on {\footnote{angel.rincon@pucv.cl}}}
\address{Instituto de F{\'i}sica, Pontificia Universidad Cat{\'o}lica de Valpara{\'i}so, Avenida Brasil 2950, Casilla 4059, Valpara{\'i}so, Chile\\}

\author{Grigoris Panotopoulos {\footnote{grigorios.panotopoulos@tecnico.ulisboa.pt}}}
\address{Centro de Astrof{\'i}sica e Gravita{\c c}{\~a}o, Departamento de F{\'i}sica, Instituto Superior T{\'e}cnico-IST, Universidade de Lisboa-UL, Av. Rovisco Pais, 1049-001 Lisboa, Portugal\\}

\author{Pedro Bargue\~no {\footnote{p.bargueno@uniandes.edu.co}}}
\address{Departamento de F\'{\i}sica,\\ Universidad de Los Andes, Cra. 1 E No 18 A-10, Bogot\'a, Colombia\\}

\begin{abstract}
In this work, starting from a spherically symmetric polytropic black hole, a rotating solution is obtained by following the Newman--Janis 
algorithm without complexification. Besides studying the
horizon, the static conditions and causality issues of the rotating solution, we obtain and discuss the shape of its shadow. Some other physical features as the Hawking temperature and emission rate of the rotating
polytropic black hole solution are also discussed.
\end{abstract}

\maketitle

\section{Introduction}\label{intro}

After the historical direct detection of gravitational waves from a black hole (BH) merger by the LIGO collaboration in 2015 \cite{ligo}, and the first image of the supermassive BH located at the centre of the giant elliptical galaxy Messier 87 (M87) by the Event Horizon Telescope (EHT) project \cite{project} only a few weeks back, there is currently a lot of ongoing research in investigating the properties of BHs in several different contexts. The LIGO direct detection provided us with the strongest evidence so far that BHs do exist in Nature and that they merge, while at the same time it offers us the tools to test strong gravitational fields using radio wave astronomy. However, the LIGO detection has provided us with no information about the horizon of the BH, which after all is its defining property. 

The gravitational field produced by a BH is so strong that nothing, not even photons, can escape. Therefore, a BH cannot be seen directly. There is, however, the possibility of seeing a dark shadow of a BH via strong gravitational lensing and photon capture at the horizon, if the BH stands between the observer and nearby light sources. The photons emitted from the source, or of the radiation emitted from an accretion flow around the event horizon of the BH, are expected to create a characteristic shadow-like image, that is a darker region over a brighter background. Indeed, recently the EHT project, a global very long baseline interferometer array observing at a wavelength of $1.3$~mm, announced and showed the first image of the supermassive BH located at the centre of M87 \cite{L1}, while the corresponding image from the centre of the Milky Way is yet to come. For instrumentation, data processing and calibration, physical origin of the shadow,
 etc, see \cite{L2,L3,L4,L5,L6}. 

The observation of the shadow probes the spacetime geometry in the vicinity of the horizon, and therefore it tests the existence and properties of the latter \cite{psaltis}. It should be noted, however, that other horizonless compact objects that possess light rings also cast shadows 
\cite{horizonless1,horizonless2,horizonless3,horizonless4,horizonless5,horizonless6,shakih2018,shakih2019}, and therefore the presence of a shadow does not by itself imply that the object is necessarily a BH. Therefore, strong lensing images and shadows offer us an exciting opportunity not only to detect the nature of a compact object, but also to test whether or not the gravitational field around a compact object is described by a rotating or non-rotating geometry. For a recent brief review on shadows see \cite{review}. 

Within the framework of Einstein's General Relativity \cite{GR} the most general BH solution is the Kerr-Newman geometry characterized by its mass, angular momentum and electric charge, see e.g. \cite{solutions}. Since, however, astrophysical BHs are expected to be electrically neutral, the most interesting cases to be considered are either the Schwarzschild \cite{SBH} or the Kerr geometry \cite{kerr}. More rotating BH solutions may be generated starting from non-rotating seed spacetimes applying the Newman-Janis algorithm (NJA), described in \cite{algorithm1,algorithm2}. 

Non--rotating solutions have been obtained in non-standard scenarios, such as polytropic BHs \cite{polytropic1,polytropic2} or BHs with quintessencial energy \cite{quint1}, to mention just a few. Over the years the shadow of the Schwarzschild geometry was considered in \cite{synge,luminet}, while the shadow cast by 
the Kerr solution was studied in \cite{bardeen} (see also \cite{monograph}). 
Shadows of Kerr BHs with scalar hair and  BH shadows in other frameworks have been considered in \cite{carlos1,carlos2} and 
\cite{Bambi:2008jg,Bambi:2010hf,study1,study2,Moffat,quint2,study3,study4,study5,study6,bobir2017,Konoplya:2019sns,shakih2019b,sudipta2019}, respectively. 
  
To explore the physics behind the so-called BH shadows, an alternative tool is provided by the well-known NJA. As already mentioned, this method allows one 
to pass from a static spherically symmetric BH solution to a rotating one. To be more precise, in the present work we will take a variation of the usual NJA, the only difference being the omission of one of the steps of the NJA, namely the complexification of coordinates \cite{quint1}. Instead of this, we will follow an ``alternate" coordinate transformation, which will be explained in the next section.

In the present work we propose to investigate the shadow of the rotating polytropic BH. The non-rotating, static, spherically symmetric geometry was obtained in \cite{polytropic1}. The metric tensor is a solution to Einstein's field equations with negative cosmological constant, and the thermodynamics of the BH precisely matches that of a polytropic gas. One one hand, as stated in \cite{polytropic2}, 
from the point of view of a possible astrophysical tests of the non-rotating polytropic solution, the so-called static radius,
which defines the equilibrium region between gravitational attraction and dark energy repulsion, would be of importance \cite{Panotopoulos:2017clp}. On the other hand, 
in the rotating case, which will be studied in the present work by using the NJA, the computation of the shadow would constitute a valuable tool in order
to confirm or refute theoretical predictions regarding the intimate structure of space and time at the strong field regime.

The plan of our work is the following. After this introduction, we briefly summarize how the NJA works in the next section. In section
\ref{ng} we study the conditions  leading to unstable null trajectories for a general parametrization of a rotating BH 
while section \ref{hter} is devoted to the study of the Hawking temperature and the emission rate of
a generic $3+1$ rotating BH. In section \ref{rotating} we construct the rotating
solution starting from a static and spherically symmetric polytropic BH as the seed geometry
and we study some features of the solution as for example horizon and static conditions, 
causality issues, BH shadow, Hawking temperature and emission rate. Finally we conclude our work in the last section. We adopt the mostly negative metric signature $(+,-,-,-)$, and we choose natural units where $c=1=G$.

\section{Newmann--Janis algorithm without complexification}\label{NJ}

In this section we review the main aspects on the NJA to generate rotating solutions introduced by M. Azreg-A\"{i}nou in Ref. \cite{azreg2014}. In this work, the author performed a modification in the algorithm with the purpose to avoid the complexification process
in the original protocol as follows. 

As usual, the starting point is a static spherically symmetric metric parametrized as
\begin{eqnarray}
ds^{2}=G(r)dt^{2}-\frac{dr^{2}}{F(r)}-H(r)(d\theta^{2}+\sin^{2}\theta d\phi^{2}).
\end{eqnarray}
The next step consists in introducing the advanced null coordinates $(u,r,\theta,\phi)$ defined by
\begin{equation}
du=dt-dr/\sqrt{FG},
\end{equation}
from where, the non--zero components of the inverse metric can be written as
$g^{\mu\nu}=l^{\mu}n^{\nu}+l^{\nu}n^{\mu}-m^{\mu}\bar{m}^{\nu}-m^{\nu}\bar{m}^{\mu}$
with
\begin{eqnarray}
l^{\mu}&=&\delta^{\mu}_{r} ,\\
n^{\mu}&=&\sqrt{F/G}\delta^{\mu}_{u}-(F/2)\delta^{\mu}_{r} ,\\
m^{\mu}&=&(\delta^{\mu}_{\theta}+\frac{i}{\sin\theta}\delta^{\mu}_{\phi})/\sqrt{2H} ,
\end{eqnarray}
and $l_{\mu}l^{\mu}=m_{\mu}m^{\mu}=n_{\mu}n^{\mu}=l_{\mu}m^{\mu}=n_{\mu}m^{\mu}=0$ and
$l_{\mu}n^{\mu}=-m_{\mu}\bar{m}^{\mu}=1$. Now, introducing the complex transformation
\begin{eqnarray}
&&r\to r+ia \cos\theta ,\\
&&u\to u-ia \cos\theta ,
\end{eqnarray}
and assuming that 
\begin{eqnarray}
&&G(r)\to A(r,\theta, a) ,\\
&&F(r)\to B(r,\theta, a) ,\\
&&H(r)\to \Psi(r,\theta, a) ,
\end{eqnarray}
we obtain
\begin{eqnarray}
l^{\mu}&=&\delta^{\mu}_{r} ,\\
n^{\mu}&=&\sqrt{B/A}\delta^{\mu}_{u}-(B/2)\delta^{\mu}_{r} ,\\
m^{\mu}&=&(\delta^{\mu}_{\theta}+ia\sin\theta(\delta^{\mu}_{u}-\delta^{\mu}_{r})+\frac{i}{\sin\theta}\delta^{\mu}_{\phi})/\sqrt{2\Psi}.
\end{eqnarray}
Using the above transformations, the line element, in the so--called rotating Eddington-–Finkelstein coordinates reads
\begin{align}\label{EF}
ds^{2}&=A du^{2}+2\sqrt{\frac{A}{B}}du dr
  +2a \sin^{2}\theta(\sqrt{\frac{A}{B}}-A)du d\phi\nonumber \\
&-2a \sin^{2}\theta\sqrt{\frac{A}{B}}dr d\phi-\Psi d\theta^{2}\nonumber\\
&-\sin^{2}\theta(\Psi + a^{2}\sin^{2}\theta(2\sqrt{\frac{A}{B}}-A))d\phi^{2}).
\end{align}
In order to write the metric (\ref{EF}) in the  Boyer–-Lindquist coordinates, we 
proceed to perform the global coordinate transformation
\begin{eqnarray}\label{bl}
du&=&dt+\lambda(r)dr ,\\
d\phi&=&d\phi+\chi(r)dr ,
\end{eqnarray}
where $\lambda$ and $\chi$ must depend on $r$ only to ensure the integrability of
Eq. (\ref{bl}). As it is well known, in the original NJA the next step in the construction of the rotating metric, consists in complexifying 
$r$. However, in order to
circumvent the complexification, Azreg-A\"{i}nou (see Ref. \cite{azreg2014}) proposed 
an ansatz for the unknown functions involved. Namely, taking
\begin{eqnarray}
\lambda&=&-\frac{(K+a^{2})}{FH+a^{2}} ,\\
   \chi&=&-\frac{a}{FH+a^{2}} ,
\end{eqnarray}
where
\begin{eqnarray}
K=\sqrt{\frac{F}{G}}H ,
\end{eqnarray}
and 
\begin{eqnarray}
A(r,\theta)&=&\frac{FH+a^{2}\cos^{2}\theta}{(K+a^{2}\cos^{2}\theta)^{2}}\Psi ,\\
B(r,\theta)&=&\frac{FH+a^{2}\cos^{2}\theta}{\Psi},
\end{eqnarray}
the metric (\ref{EF}) takes the Kerr--like form
\begin{eqnarray}\label{blkf}
ds^{2}&=&\frac{\Psi}{\rho^{2}}\bigg(\frac{\Delta}{\rho^{2}}(dt-a \sin^{2}\theta d\phi)^{2}-\frac{\rho^{2}}{\Delta}dr^{2}-\rho^{2}d\theta^{2}\nonumber\\
&&-\frac{\sin^{2}\theta}{\rho^{2}}(adt-(K+a^{2})d\phi)^{2})\bigg),
\end{eqnarray}
with
\begin{eqnarray}
\rho^{2}&=&K+a^{2}\cos^{2}\theta ,
\end{eqnarray}
where $a=J/M$, with $M,J$ being the mass and the rotation speed, respectively, of the black hole.

At this point some comments are in order. First, note that the function $\Psi(r,\theta,a)$ remains unknown but it must satisfy the following differential equation
\begin{eqnarray}\label{deq}
&&(K+a^{2}y^{2})^{2}(3\Psi_{,r}\Psi_{,y^{2}}-2\Psi\Psi_{r,y^{2}})=3a^{2}K_{,r}\Psi^{2}
\end{eqnarray}
which corresponds to imposing that the Einstein tensor satisfies $G_{r\theta}=0$. Second, it can be shown (see appendix A in Ref.\cite{azreg2014}) that the metric (\ref{blkf}) satisfies Einstein's field equations $G_{\mu\nu}=8\pi T_{\mu\nu}$ with the source given by
\begin{eqnarray}
T^{\mu\nu}=\epsilon e^{\mu}_{t}e^{\nu}_{t}+p_{r}e^{\mu}_{r}e^{\nu}_{r}+p_{\theta}e^{\mu}_{\theta}e^{\nu}_{\theta}+p_{\phi}e^{\mu}_{\phi}e^{\nu}_{\phi},
\end{eqnarray}
with 
\begin{eqnarray}
e^{\mu}_{t}&=&\frac{(r^{2}+a^{2},0,0,a)}{\sqrt{\rho^{2}\Delta}} ,\\
e^{\mu}_{r}&=&\frac{\sqrt{\Delta}(0,1,0,0)}{\sqrt{\rho^{2}}} ,\\
e^{\mu}_{\theta}&=&\frac{(0,0,1,0)}{\sqrt{\rho^{2}}} ,\\
e^{\mu}_{\phi}&=&-\frac{(a\sin^{2}\theta,0,0,1)}{\sqrt{\rho^{2}}\sin\theta}.
\end{eqnarray}
Even more, in order to ensure the consistency of Einstein's field equations, the unknown 
$\Psi$ must satisfy another constraint, namely
\begin{eqnarray}
&&(K_{,r}^{2}+K(2-K_{,rr})-a^{2}y^{2}(2+K_{,rr}))\Psi , \nonumber\\
&&+(K+a^{2}y^{})(4y^{2}\Psi_{y^{2}}-K_{,r}\Psi_{,r})=0 .
\end{eqnarray}
Finally, the above expressions can be simplified in the particular case $G=F$ and $H=r^{2}$. Indeed, in this case it can be shown that one solution of 
Eq. (\ref{deq}) is given by
\begin{eqnarray}
\Psi=r^{2}+a^{2}\cos^{2}\theta ,
\end{eqnarray}
and the metric (\ref{blkf}) takes the form
\begin{eqnarray}\label{blsm}
ds^{2}&=&\left(1-\frac{2f}{\rho^{2}}\right)dt^{2}-\frac{\rho^{2}}{\Delta}dr^{2}
+\frac{4af \sin^{2}\theta}{\rho^{2}}dtd\phi\nonumber\\
&&-\rho^{2}d\theta^{2}-\frac{\Sigma \sin^{2}\theta}{\rho^{2}}d\phi^{2} ,
\end{eqnarray}
with
\begin{eqnarray}
\rho^{2}&=&r^{2}+a^{2}\cos^{2}\theta ,\\
2f & = & r^{2}(1-F) ,\\
\Delta & = &-2f+a^{2}+ r^2 , \\
\Sigma & = & (r^{2}+a^{2})^{2}-a^{2} \Delta \sin^{2} \theta.
\end{eqnarray}
It is easy to verify that when $a=0$ we recover the non rotating black hole solution.

\section{Null geodesics around the rotating black hole}\label{ng}

In this section we implement the standard Hamilton--Jacobi formalism to separate the null geodesic equations in the rotating space--time. Our main goal here is to obtain the celestial coordinates parametrized with the radius of the unstable null orbits and to study the shadow of a generic rotating solution. 

Let us start with the Hamilton--Jacobi equations \cite{carter}
\begin{eqnarray}\label{jac}
\frac{\partial S}{\partial\tau}=\frac{1}{2}g^{\mu\nu}\partial_{\mu}S\partial_{\nu}S ,
\end{eqnarray}
where $\tau$ is the proper time and $S$ is the Jacobi action. As usual, if we assume that
Eq. (\ref{jac}) has separable solutions, the action takes the form
\begin{eqnarray}\label{separable}
S=-Et+\Phi\phi+S_{r}(r)+S_{\theta}(\theta) ,
\end{eqnarray}
where $E$ and $\Phi$ are the conserved energy and angular momentum respectively. Now, replacing 
(\ref{separable}) in (\ref{jac}) we obtain that
\begin{eqnarray}
S_{r} & = & \int\limits^{r}\frac{\sqrt{R(r)}}{\Delta}dr ,\\
S_{\theta} & = & \int\limits^{\theta}\sqrt{\Theta(\theta)}d\theta,
\end{eqnarray}
where
\begin{align}
R(r)&=((r^{2}+a^{2})E-a\Phi)^{2}-\Delta(Q+(\Phi-aE)^{2}) ,\\
\Theta(\theta)&=Q-(\Phi^{2}\csc^{2}\theta-a^{2}E^{2})\cos^{2}\theta ,
\end{align}
where $Q$  is the so--called Carter constant. As it is well-known, the unstable photon orbits in the rotating
space--time must satisfy the constraints $R=0$ and $R'=0$ which lead to
\begin{align}
&\left(a^2-a \xi +r^2\right)^2-\left(a^2+r^2 F\right) \left((a-\xi )^2+\eta \right)=0 , \\
&4 \left(a^2-a \xi +r^2\right)-\left((a-\xi )^2+\eta \right) \left(r F'+2 F\right)=0,
\end{align}
where $\xi=\Phi/E$ and $\eta=Q/E^{2}$ correspond to the impact parameters. Hence, 
\begin{eqnarray}
\xi&=&-\frac{4 \left(a^2+r^2 F\right)}{a(r F'+2 F)}+a+\frac{r^2}{a}\label{chi}, \\
\eta&=&\frac{r^3 \left(8 a^2 F'-r \left(r F'-2 F\right)^2\right)}{a^2 \left(r F'+2 F\right)^2}\label{eta}.
\end{eqnarray}
It is worth mentioning that, in the above expressions, $r$ corresponds to the radius of the unstable null orbits.

Now, the apparent shape of the shadow is obtained by using the celestial coordinates which are defined as \cite{vazquez}
\begin{eqnarray}
\alpha&=&\lim\limits_{r_{0}\to \infty}\left(-r_{0}^{2}\sin\theta_{0}\frac{d\phi}{dr}\bigg|_{(r_{0,\theta_{0}})}\right), \\
\beta&=&\lim\limits_{r_{0}\to\infty}\left(r_{0}^{2}\frac{d\theta}{dr}\bigg|_{(r_{0},\theta_{0})}\right),
\end{eqnarray}
where $(r_{0},\theta_{0})$ correspond to the coordinates of the observer. Finally, the 
after calculating the limit in the above expressions, the celestial coordinates read
\begin{eqnarray}
\alpha&=&-\frac{\xi}{\sin\theta_{0}}\label{alpha} ,\\
\beta &=&\pm\sqrt{\eta+a^{2}\cos\theta^{2}_{0}-\chi^{2}\cot^{2}\theta_{0}}\label{beta} \ ,
\end{eqnarray}
and the shadow corresponds to the parametric curve of $\alpha$ and $\beta$ with $r$ as a parameter.

\section{Hawking temperature and emission rate of a rotating black hole}\label{hter}

In Ref. \cite{ma}, the author derived the Hawking temperature of a general four--dimensional rotating BH using the null--geodesic tunneling method developed by Parikh and Wilczek \cite{parikh1,parikh2,parikh3}.
In this section we shall summarize the main result obtained in \cite{ma}. 

First of all, the radiated particles are considered as $s$-waves because for an observer at infinity 
the radiation is spherically symmetric whether the BH is rotating or not. Now, the tunnelling rate of a $s$-wave from the
inside to the outside of the BH is given by
\begin{eqnarray}\label{gamma1}
\Gamma=\Gamma_{0}e^{-2Im\mathcal{I}},
\end{eqnarray}
where $\mathcal{I}$ is the action of the tunnelling particle and $\Gamma_{0}$ the normalization factor. Moreover, as it is well-known, the emission rate satisfies
\begin{eqnarray}\label{gamma2}
\Gamma=\Gamma_{0}e^{-\beta E},
\end{eqnarray}
where $E$ is the energy of the emitted particle, $\beta=2\pi/\kappa$ and $\kappa$ is the surface gravity of the horizon. Now, from
(\ref{gamma1}) and (\ref{gamma2}) 
we arrive to
\begin{eqnarray}\label{sg}
\kappa=\frac{\pi E}{Im\mathcal{I}},
\end{eqnarray}
from where the Hawking temperature can be derived form the standard relation 
\begin{eqnarray}\label{th}
T_{H}=\frac{\kappa}{2\pi}.
\end{eqnarray}

In order to obtain an expression of the Hawking temperature in terms of the components of the metric we proceed to consider a generic rotating line element of the form
\begin{eqnarray}
ds^{2}&=&-g_{tt}dt^{2}+g_{rr}dr^{2}+g_{\theta\theta}dr^{2}+g_{\theta\theta}d\theta^{2}\nonumber\\  
            &&+g_{\phi\phi}d\phi^{2}-2g_{t\phi}dtd\phi.
\end{eqnarray}
It can be demonstrated that (see section II of Ref. \cite{ma} for details)
\begin{eqnarray}\label{im}
Im\mathcal{I}=\frac{2\pi E}{\sqrt{G'_{tt}(r_{+},\theta_{0})g^{rr'}(r_{+},\theta_{0})}},
\end{eqnarray}
where the prime stands for derivative respect to the radial coordinate, $r_{+}$ is the horizon
radius and
\begin{eqnarray}
G_{tt}=g_{tt}+2g_{t\phi}\Omega_{+}-g_{\phi\phi}\Omega_{+}^{2} ,
\end{eqnarray}
with $\Omega_{+}$ the angular velocity of the event horizon, which is a constant defined by
\begin{eqnarray}
\Omega_{+}=\frac{g_{t\phi}}{g_{\phi\phi}}\bigg|_{r=r_{+}}.
\end{eqnarray}
Finally, replacing (\ref{im}) in (\ref{sg}) an using (\ref{th}), the Hawing temperature takes the form
\begin{eqnarray}
T_{H}=\frac{\sqrt{G'_{tt}(r_{+},\theta_{0})g^{rr'}(r_{+},\theta_{0})}}{4\pi} ,
\end{eqnarray}
which can be written alternatively as
\begin{eqnarray}\label{Temp}
T_{H}=\frac{1}{4\pi}\lim\limits_{r\to r_{+}}\frac{\partial_{r}G_{tt}}{\sqrt{G_{tt}g_{rr}}}.
\end{eqnarray}

Please, notice that the Eq.~\eqref{Temp} is a generalization of the standard formula for the Hawking temperature. In particular, when the angular velocity is taken to be zero, we recover the well--known formula, namely:
\begin{align}
T_{H}=\frac{1}{4\pi} \lim\limits_{r\to r_{+}}\frac{\partial_{r}g_{tt}}{\sqrt{g_{tt}g_{rr}}} .
\end{align}

In what follows, we shall study the emission rate of a rotating BH. As it is claimed in references \cite{amir2016,papnoi2014}, 
the BH shadow corresponds to its high energy absorption cross section for the 
observer located at infinity. On the other hand, it is well known that for a BH endowed with a photon sphere, the cross section
has a limiting value, namely, $\sigma_{lim}$, which coincides with the geometrical cross section of this photon sphere \cite{misner}. Now, for a spherically symmetric BH it can be demonstrated that \cite{shao2013}, $\sigma_{lim}\approx \pi R_{s}^{2}$ with
$R_{s}$ the radius of the photon sphere defined by (see \cite{amir2016})
\begin{eqnarray}\label{radi}
R_{s}=\frac{(\alpha_{t}-\alpha_{r})^{2}+\beta_{t}^{2}}{2|\alpha_{t}-\alpha_{r}|}.
\end{eqnarray}
In the previous expression, the quantities $\alpha_{t},\alpha_{r},\beta_{t}$ correspond to particular values of 
the celestial coordinates (see  Eqs. (\ref{alpha}) and (\ref{beta})). In particular, $\alpha_{r}$ corresponds to the
most right value of $\alpha$ and the pair $(\alpha_{t},\beta_{t})$ stands for coordinates of the top of the shadow \cite{hioki2009}. 
With all the quantities defined in this section, the emission rate \cite{spectrum,wei2013,amir2016,Panotopoulos:2016wuu,Panotopoulos:2017yoe,Destounis:2018utr,Panotopoulos:2018pvu,Rincon:2018ktz} can be calculated by
\begin{eqnarray}\label{er}
\frac{d^{2}E}{d\omega dt } = \frac{\sigma_{lim}}{e^{\omega / T_H}-1} \frac{\omega^{3}}{2 \pi^2},
\end{eqnarray}
where $\omega$ represents the frequency of photons. In the next section we obtain the Hawing temperature and the emission rate for a rotating polytropic BH.

\section{Rotating polytropic black hole solution}\label{rotating}

In this section we construct the rotating BH solution from a static and spherically symmetric 
polytropic BH. Then we shall study some of its physical properties as well as the BH shadow and its emission. 

As a starting point, we consider the polytropic BH solution obtained in \cite{polytropic1}
\begin{align}\label{metric}
&ds^{2}=\mathcal{F}dt^{2}-\mathcal{F}^{-1} d r^{2}-r^{2}d \Omega^{2},
\end{align}
where $\mathcal{F}=\left(\frac{r^2}{L^2}-\frac{2 M}{r}\right)$, $L^{2}=-3/\Lambda$, with $\Lambda$ being the cosmological constant, and 
$d \Omega^2 = d \theta^{2} + \sin^{2} \theta d \phi^{2}$ is the line element of the unit two-dimensional sphere. It is worth mentioning that $\mathcal{F}$ in (\ref{metric}) has the same form
of the black string solution found in \cite{lemos,cai} in the context
of plane symmetric solutions of Einstein's equations \cite{solutions}.

Now, using $F=G=\frac{r^2}{L^2}-\frac{2 M}{r}$, the
line element (\ref{blsm}) reads
\begin{eqnarray}
ds^{2}&=&\left(1-\frac{r^{2}}{\rho ^2}
\left(1-\frac{r^2}{L^2}+\frac{2 M}{r}\right)\right)dt^2 
-\frac{\rho }{\Delta }dr^2  - \rho ^2 d\theta^2\nonumber\\
&&+\frac{2 a r^{2}  \sin ^2\theta }{\rho ^2} \left(1-\frac{r^2}{L^2}+\frac{2 M}{r}\right)dtd\phi\nonumber\\
&&-\frac{\Sigma  \sin ^2\theta }{\rho ^2}d\phi^2 ,
\end{eqnarray}
with 
\begin{eqnarray}
\Delta=a^2+r^2 \left(\frac{r^2}{L^2}-\frac{2 M}{r}\right).
\end{eqnarray}
At this point some comments are in order. First, the horizons are solutions of $\Delta(r_{\pm})=0$
which, in this particular case, corresponds to
\begin{eqnarray}\label{hrc}
a^2+r_{\pm}^2 \left(\frac{r_{\pm}^2}{L^2}-\frac{2 M}{r_{\pm}}\right)=0.
\end{eqnarray}
It is worth mentioning that from the horizon condition we can obtain a bound on the spin parameter, namely $a/M$. In this case, the 
allowed values are constrained by
\begin{eqnarray}
\frac{a}{M}<\frac{3^{1/2}}{2^{2/3}}\left(\frac{L}{M}\right)^{1/3}.
\end{eqnarray}
Note that above result differs from the obtained in the case of the Kerr solution, where the constraint is given by $a/M<1$.
Second, the static limit, namely, the surface from where observers can remain static, corresponds to
$g_{tt}=0$. More precisely,
\begin{eqnarray}
\frac{a^2 \cos ^2(\theta )+r_{st}^2 \left(\frac{r_{st}^2}{L^2}-\frac{2 M}{r_{st}}\right)}{a^2 \cos ^2(\theta )+r_{st}^2}=0.
\end{eqnarray}
Note that, the event horizon, $r_{+}$, coincides with
the static radius $r_{st}$ at the poles $\theta=0$ and $\theta=\pi$ as in the Kerr solution. Third, note that causality violation and closed time--like 
curves are possible if $g_{\phi\phi}>0$. To
be more precise, the condition is
\begin{eqnarray}
-\frac{\Sigma  \sin ^2(\theta )}{\rho ^2}>0,
\end{eqnarray}
from where, given that $\sin^{2}(\theta)/\rho ^2$ is positive, the sign of
$\Sigma$ plays a crucial role in the analysis. What is more, the condition to avoid causality violation and closed time--like curves is to impose $\Sigma>0$. In the particular case of the rotating polytropic BH, the condition on $\Sigma$ reads
\begin{eqnarray}
a^2 \sin ^2(\theta ) \left(-a^2-\frac{r^4}{L^2}+2 M r\right)+\left(a^2+r^2\right)^2>0,
\end{eqnarray}
It is worth mentioning that, in contrast to Kerr solution, the 
causality issues can be avoided for particular choices of the parameters. For example, taking
$a=L$, the above condition reduces to
\begin{eqnarray}
L^4+2 L^2 M r \sin ^2\theta+2 L^2 r^2+r^4 \left(1 - \sin ^2\theta\right)>L^4,
\end{eqnarray}
which is trivially satisfied given that all the coefficients involved in the left hand side are
positive. 

Now, we shall focus our attention in the construction of the BH shadow. Replacing the metric function 
$F(r)=-\frac{2M}{r}+\frac{r^{2}}{L^{2}}$ in
Eqs. (\ref{chi}) and (\ref{eta}), the impact parameters $(\xi,\eta)$ associated to the unstable null geodesics around the rotating BH
are given by
\begin{eqnarray}
\xi&=&\frac{L^2 r \left(2 a^2-3 M r\right)}{a \left(L^2 M-2 r^3\right)}+a ,\\
\eta&=&\frac{L^2 r^3 \left(4 a^2 \left(L^2 M+r^3\right)-9 L^2 M^2 r\right)}{a^2 \left(L^2 M-2 r^3\right)^2},
\end{eqnarray}
from where, assuming $\theta_{0}=\pi/2$, the celestial coordinates read
\begin{eqnarray}
\alpha &=&-a+ \frac{L^2 r \left(3 M r-2 a^2\right)}{a \left(L^2 M-2 r^3\right)} ,\\
\beta&=&\pm \sqrt{\frac{L^2 r^3 \left(4 a^2 \left(L^2 M+r^3\right)-9 L^2 M^2 r\right)}{a^2 \left(L^2 M-2 r^3\right)^2}}.
\end{eqnarray}

\begin{figure*}[h!]
\centering
\includegraphics[scale=0.5]{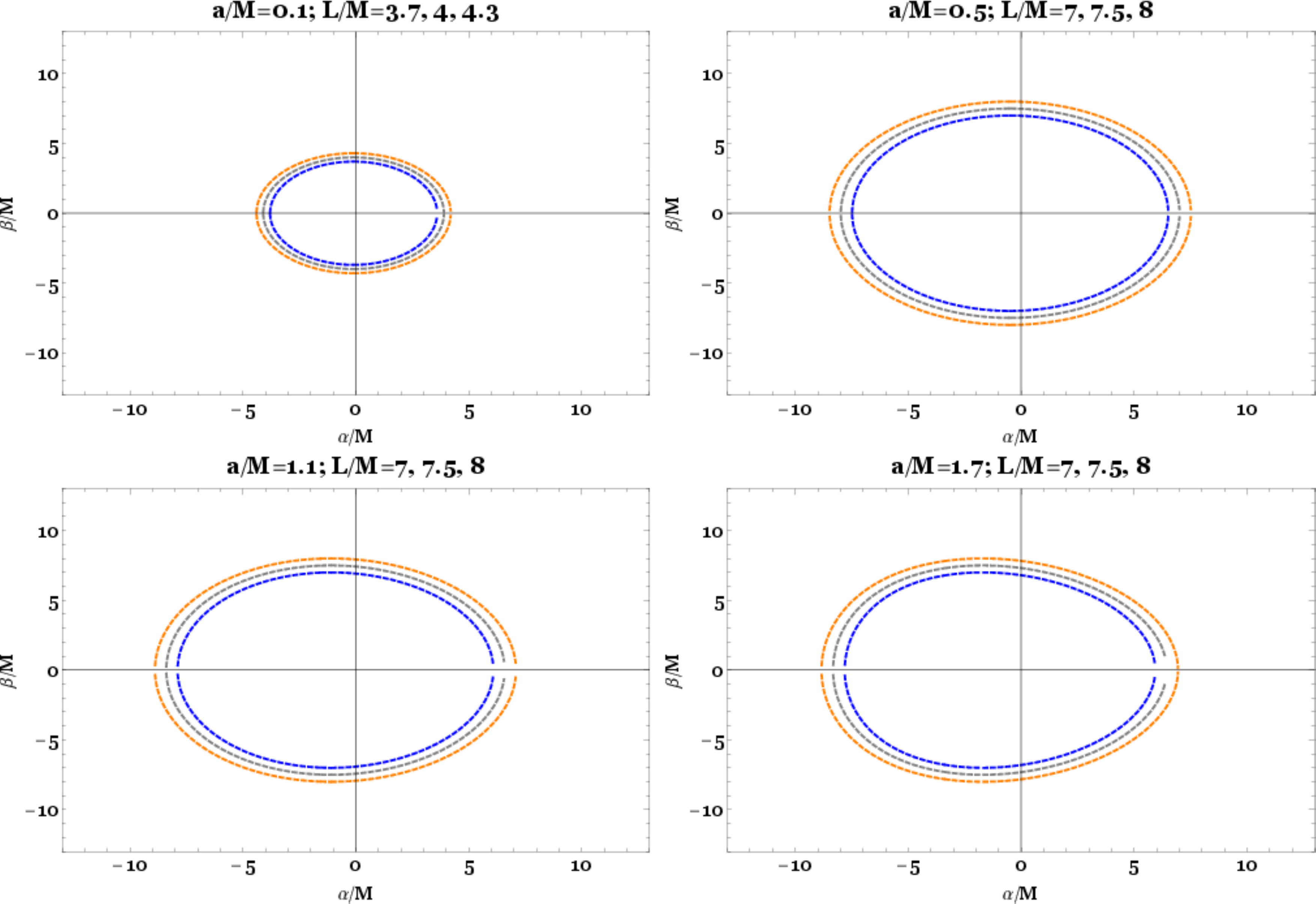}
\caption{\label{fig1} 
Silhouette of the shadow cast by the rotating polytropic BH for $a/M=0.1$ (left upper panel)
with $L/M=3.7$ (blue line) $L/M=4$ (gray line) $L/M=4.3$ (orange line) and $a/M=0.5$ (right upper panel),
$a/M=1.1$ (left down panel) and $a/M=1.7$ (right down panel) with
$L/M=7$ (blue line), $L/M=7.5$ (gray line) and $L/M=8$ (orange line).
}
\end{figure*}

In figure \ref{fig1} the shadow of the rotating BH is shown for different values of $a/M$ and $L/M$.
Note that the shadow undergoes a deformation as 
$a/M$ increases. Indeed, for $a/M=1.5$ the silhouette losses its symmetry and
looks like an oval instead of an ellipse. 

%%%%%%%%%%%%%%%%%%%%%%%%%%%%%%%%%%%%%%%%%%%%%%%%%%%%%%%%%%%%%%%%%%%%%%

%%%%%%%%%%%%%%%%%%%%%%%%%%%%%%%%%%%%%%%%%%%%%%%%%%%%%%%%%%%%%%%%%%%%%%%%%%

\begin{figure}[h!]
\centering
\includegraphics[scale=0.5]{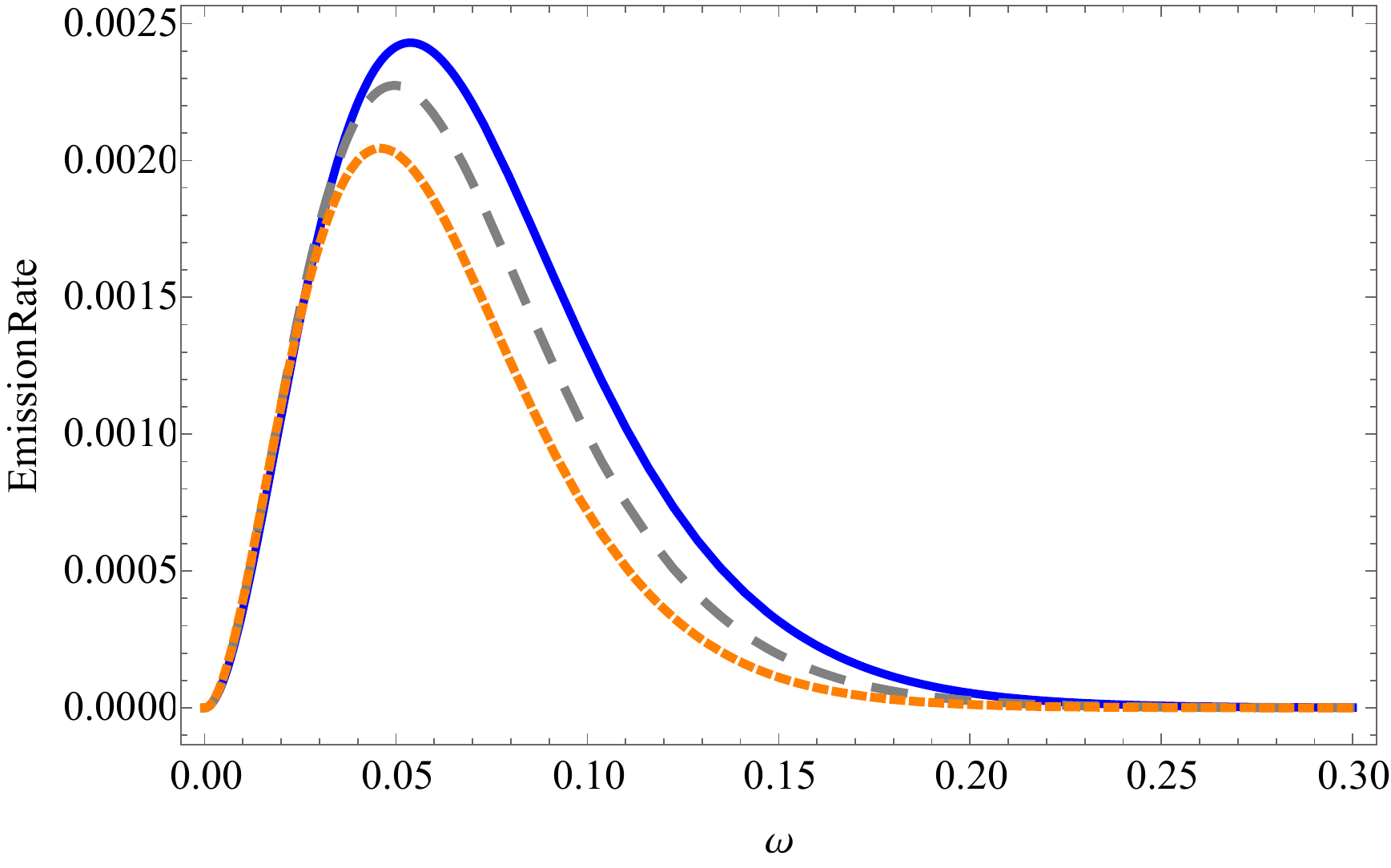}
\caption{\label{figemission} 
Emission rate of the rotating polytropic BH for $a/M=0.1$
and $L/M=7$ (solid blue line), $L/M=7.5$ (dashed gray line) and $L/M=8$ (dotted orange line)
}
\end{figure}

The hawking temperature for this rotating BH is given by
\begin{eqnarray}\label{htp}
T_{H}=\frac{-a^2 \left(L^2 (M+r_{+})-2 r_{+}^3\right)+L^2 M r_{+}^2+r_{+}^5}{2 \pi  L^2 \left(a^2+r_{+}^2\right)^2},
\end{eqnarray}
where $r_{+}$ is the event horizon radius obtained from the condition (\ref{hrc}), which reads
\begin{align}
r_{+}&=\frac{\sqrt{-\frac{2 \sqrt[3]{6} a^2 L^2}{d(a,M,L)}+\frac{6 L^2 M}{\sqrt{\frac{\sqrt[3]{6} a^2 L^2}{d(a,M,L)}+\frac{1}{2} d(a,M,L)}}-d(a,M,L)}}{2^{5/6} \sqrt[3]{3}}\nonumber\\
&+\frac{\sqrt{\frac{2 \sqrt[3]{6} a^2 L^2}{d(a,M,L)}+d(a,M,L)}}{2^{5/6} \sqrt[3]{3}},
\end{align}
with
\begin{eqnarray}
d(a,M,L)=\sqrt[3]{\sqrt{3} \sqrt{27 L^8 M^4-16 a^6 L^6}+9 L^4 M^2}.
\end{eqnarray}
Now, for particular values of the parameters $(a,L,M)$, the Hawking temperature can be computed and $R_{s}$ can be obtained from 
Eq. (\ref{radi}). Finally, replacing (\ref{htp}) and (\ref{radi}) in
(\ref{er}), we can study the behaviour of the emission rate as a function of the frequency $\omega$. The emission rate profile is depicted in 
figure \ref{figemission}, where it can be shown that it decreases with an increasing value of $L/M$.

%%%%%%%%%%%%%%%%%%%%%%%%%%%%%%%%%%%%%%%%%%%%%%%%%%%%%%%%%%%%%%%%%%%%%%%%%%%%%

%%%%%%%%%%%%%%%%%%%%%%%%%%%%%%%%%%%%%%%%%%%%%%%%%%%%%%%%%%%%%%%%%%%%%%%%%%%%%%%

\section{Conclusions}\label{remarks}

In this work we have reviewed the main aspects related to the Newman--Janis algorithm without complexification,
the construction of unstable null orbits and the computation of the Hawking temperature and emission rate for general rotating black holes and we have used these tools to construct
a rotating polytropic black hole solution. It is worth mentioning that, to the best of our knowledge, the construction of such a rotating solution has not been considered before. Besides, we have studied some physical properties as the position of the horizons and the static limit 
which defines the ergosphere as well as the causality condition. We have obtained that, in contrast 
to the Kerr solution, the causality issues can be avoided for certain choice of the black hole parameters, namely $a$, $M$
and $L$. Additionally, we have demonstrated that, in contrast to the Kerr black hole, the bound on the
spin parameter $a/M$ can be grater that one because the condition of the appearance
of horizons entails that this bound is proportional to $L/M$. We have also analysed the shadow of the rotating polytropic 
solution finding that its shape can change while increasing $L/M$. As a final result, we have studied the emission rate, showing that it increases for large values of $L/M$.

Finally, it would be interesting to investigate how the properties of the solution here obtained are modified in light of the so called 
scale--dependent scenario. In such cases, the coupling constants acquire a dependence on the scale with a free parameter which encodes the quantum features 
(which is called scale--dependent parameter). 
Thus, given that the black hole shadows give observational evidence for black holes, we could establish 
bounds on the aforementioned parameter (see 
\cite{Koch:2016uso,Rincon:2017ypd,Rincon:2017goj,Rincon:2017ayr,Contreras:2017eza,Rincon:2018sgd,Hernandez-Arboleda:2018qdo,Contreras:2018dhs,Rincon:2018dsq,Contreras:2018gct,
Canales:2018tbn,Rincon:2019cix,Rincon:2019zxk,Contreras:2019fwu,Contreras:2018gpl,Contreras:2018swc}and references therein). We hope to be able to address this issue in a future work.

%%%%%%%%%%%%%%%%%%%%%%%%%%%%%%%%%%%%%%%%%%%%%%%%%%%%%%%%%%%%%%%%%%%%%%%%%%%%%%%%%%%%%%%%%%%%%%%

\section*{Acknowlegements}

The author \'A.~R. acknowledges DI-VRIEA for financial support through Proyecto Postdoctorado 2019 VRIEA-PUCV. The author G.~P. thanks the Funda\c c\~ao para a Ci\^encia e Tecnologia (FCT), Portugal, for the financial support to the Center for Astrophysics and Gravitation-CENTRA, Instituto Superior T{\'e}cnico,  Universidade de Lisboa, through the Grant No. UID/FIS/00099/2013. The author P. B. was
supported by the Faculty of Science and Vicerrector\'{\i}a de Investigaciones of Universidad de los Andes, Bogot\'a, Colombia, 
under the grant number INV-2018-50-1378.

%%%%%%%%%%%%%%%%%%%%%%%%%%%%%%%%%%%%%%%%%%%%%%%%%%%%%%%%%%%%%%%%%%%%%%%%%%%%%%%%%%%%%%%%%%%%%%%%


\begin{thebibliography}{99}
\bibitem{ligo} B.~P.~Abbott {\it et al.} [LIGO Scientific and Virgo Collaborations],
  %``Observation of Gravitational Waves from a Binary Black Hole Merger,''
  Phys.\ Rev.\ Lett.\  {\bf 116} (2016) no.6,  061102
[arXiv:1602.03837 [gr-qc]].

\bibitem{project} 
\url{https://eventhorizontelescope.org}

\bibitem{L1} K.~Akiyama {\it et al.} [Event Horizon Telescope Collaboration],
  %``First M87 Event Horizon Telescope Results. I. The Shadow of the Supermassive Black Hole,''
  Astrophys.\ J.\  {\bf 875} (2019) no.1,  L1.

\bibitem{L2} K.~Akiyama {\it et al.} [Event Horizon Telescope Collaboration],
  %``First M87 Event Horizon Telescope Results. II. Array and Instrumentation,''
  Astrophys.\ J.\  {\bf 875} (2019) no.1,  L2.

\bibitem{L3} K.~Akiyama {\it et al.} [Event Horizon Telescope Collaboration],
  %``First M87 Event Horizon Telescope Results. III. Data Processing and Calibration,''
  Astrophys.\ J.\  {\bf 875} (2019) no.1,  L3.

\bibitem{L4} K.~Akiyama {\it et al.} [Event Horizon Telescope Collaboration],
  %``First M87 Event Horizon Telescope Results. IV. Imaging the Central Supermassive Black Hole,''
  Astrophys.\ J.\  {\bf 875} (2019) no.1,  L4.

\bibitem{L5} K.~Akiyama {\it et al.} [Event Horizon Telescope Collaboration],
  %``First M87 Event Horizon Telescope Results. V. Physical Origin of the Asymmetric Ring,''
  Astrophys.\ J.\  {\bf 875} (2019) no.1,  L5.

\bibitem{L6} K.~Akiyama {\it et al.} [Event Horizon Telescope Collaboration],
  %``First M87 Event Horizon Telescope Results. VI. The Shadow and Mass of the Central Black Hole,''
  Astrophys.\ J.\  {\bf 875} (2019) no.1,  L6.
  
\bibitem{psaltis} A.~E.~Broderick, T.~Johannsen, A.~Loeb and D.~Psaltis,
  %``Testing the No-Hair Theorem with Event Horizon Telescope Observations of Sagittarius A*,''
  Astrophys.\ J.\  {\bf 784} (2014) 7
%  doi:10.1088/0004-637X/784/1/7
  [arXiv:1311.5564 [astro-ph.HE]].
  
\bibitem{horizonless1} P.~G.~Nedkova, V.~K.~Tinchev and S.~S.~Yazadjiev,
  %``Shadow of a rotating traversable wormhole,''
  Phys.\ Rev.\ D {\bf 88} (2013) no.12,  124019
%  doi:10.1103/PhysRevD.88.124019
  [arXiv:1307.7647 [gr-qc]].
  
\bibitem{horizonless2} F.~H.~Vincent, Z.~Meliani, P.~Grandclement, E.~Gourgoulhon and O.~Straub,
  %``Imaging a boson star at the Galactic center,''
  Class.\ Quant.\ Grav.\  {\bf 33} (2016) no.10,  105015
%  doi:10.1088/0264-9381/33/10/105015
  [arXiv:1510.04170 [gr-qc]].
  
\bibitem{horizonless3} T.~Ohgami and N.~Sakai,
  %``Wormhole shadows,''
  Phys.\ Rev.\ D {\bf 91} (2015) no.12,  124020
%  doi:10.1103/PhysRevD.91.124020
  [arXiv:1704.07065 [gr-qc]].

\bibitem{horizonless4} T.~Ohgami and N.~Sakai,
  %``Wormhole shadows in rotating dust,''
  Phys.\ Rev.\ D {\bf 94} (2016) no.6,  064071
%  doi:10.1103/PhysRevD.94.064071
  [arXiv:1704.07093 [gr-qc]].

\bibitem{horizonless5} G.~Gyulchev, P.~Nedkova, V.~Tinchev and S.~Yazadjiev,
  %``On the shadow of rotating traversable wormholes,''
  Eur.\ Phys.\ J.\ C {\bf 78} (2018) no.7,  544
%  doi:10.1140/epjc/s10052-018-6012-9
  [arXiv:1805.11591 [gr-qc]].


\bibitem{horizonless6} A.~B.~Abdikamalov, A.~A.~Abdujabbarov, D.~Ayzenberg, D.~Malafarina, C.~Bambi and B.~Ahmedov,
  %``A black hole mimicker hiding in the shadow: Optical properties of the $\gamma$ metric,''
  arXiv:1904.06207 [gr-qc].  

\bibitem{shakih2018} R. Shaikh, Phys. Rev. D {\bf 98}, 024044 (2018).

\bibitem{shakih2019}  R. Shaikh, P. Kocherlakota, R. Narayan, and P. S. Joshi, Mon. Not. R.
Astron. Soc. {\bf 482}, 52 (2019).
  

\bibitem{review} P.~V.~P.~Cunha and C.~A.~R.~Herdeiro,
  %``Shadows and strong gravitational lensing: a brief review,''
  Gen.\ Rel.\ Grav.\  {\bf 50} (2018) no.4,  42
%  doi:10.1007/s10714-018-2361-9
  [arXiv:1801.00860 [gr-qc]].
  
\bibitem{GR} A.~Einstein, 
  % "The Foundation of the General Theory of Relativity," 
Annalen Phys. 49 (1916) 769–822.

\bibitem{solutions} H.~Stephani, D.~Kramers, M.~A.~H.~MacCallum, C.~Hoenselaers, C.~Herlt, \textit{Exact solutions of Einstein's field equations}, Cambridge University Press (Cambridge, United Kingdom, 2003)
  
\bibitem{SBH} K.~Schwarzschild,
  %``On the gravitational field of a mass point according to Einstein's theory,''
Sitzungsber.\ Preuss.\ Akad.\ Wiss.\ Berlin (Math.\ Phys.\ ) {\bf 1916} (1916) 189 [physics/9905030].

\bibitem{kerr} R.~P.~Kerr,
  %``Gravitational field of a spinning mass as an example of algebraically special metrics,''
  Phys.\ Rev.\ Lett.\  {\bf 11} (1963) 237.
  
\bibitem{algorithm1} E.~T.~Newman, R.~Couch, K.~Chinnapared, A.~Exton, A.~Prakash and R.~Torrence,
  %``Metric of a Rotating, Charged Mass,''
  J.\ Math.\ Phys.\  {\bf 6} (1965) 918.
  
\bibitem{algorithm2} E.~T.~Newman and A.~I.~Janis,
  %``Note on the Kerr spinning particle metric,''
  J.\ Math.\ Phys.\  {\bf 6} (1965) 915. 
 
 
\bibitem{polytropic1} M.~Setare and H.~Adami, Phys. \ Rev. \ D {\bf 91}, 084014 (2015).

\bibitem{polytropic2} E.~Contreras, \'A.~Rinc\'on, B.~Koch and P.~Bargue\~no,
  %``Scale-dependent polytropic black hole,''
  Eur.\ Phys.\ J.\ C {\bf 78} (2018) no.3,  246
%  doi:10.1140/epjc/s10052-018-5709-0
  [arXiv:1803.03255 [gr-qc]].
  
\bibitem{quint1} B.~Toshmatov, Z.~Stuchl\'ik and B.~Ahmedov,
  %``Rotating black hole solutions with quintessential energy,''
  Eur.\ Phys.\ J.\ Plus {\bf 132} (2017) no.2,  98
%  doi:10.1140/epjp/i2017-11373-4
  [arXiv:1512.01498 [gr-qc]].     

\bibitem{synge} J.~L.~Synge,
  %``The Escape of Photons from Gravitationally Intense Stars,''
  Mon.\ Not.\ Roy.\ Astron.\ Soc.\  {\bf 131} (1966) no.3,  463.
 
\bibitem{luminet} J.-P.~Luminet,
  %``Image of a spherical black hole with thin accretion disk,''
  Astron.\ Astrophys.\  {\bf 75} (1979) 228.

\bibitem{bardeen} J.~M.~Bardeen, in \textit{Black Holes} (Les Astres Occlus), 
edited by C.~Dewitt and B.~S.~Dewitt (Gordon and Breach, New York, 1973), pp. 215-239.

\bibitem{monograph} S.~Chandrasekhar, \textit{The mathematical theory of black holes}, OXFORD, UK: CLARENDON (1985) 646 P.

\bibitem{carlos1} P.~V.~P.~Cunha, C.~A.~R.~Herdeiro, E.~Radu and H.~F.~Runarsson,
  %``Shadows of Kerr black holes with scalar hair,''
  Phys.\ Rev.\ Lett.\  {\bf 115} (2015) no.21,  211102
%  doi:10.1103/PhysRevLett.115.211102
  [arXiv:1509.00021 [gr-qc]].

\bibitem{carlos2} P.~V.~P.~Cunha, C.~A.~R.~Herdeiro, E.~Radu and H.~F.~Runarsson,
  %``Shadows of Kerr black holes with and without scalar hair,''
  Int.\ J.\ Mod.\ Phys.\ D {\bf 25} (2016) no.09,  1641021
%  doi:10.1142/S0218271816410212
  [arXiv:1605.08293 [gr-qc]].

\bibitem{Bambi:2008jg}
  C.~Bambi and K.~Freese,
  %``Apparent shape of super-spinning black holes,''
  Phys.\ Rev.\ D {\bf 79}, 043002 (2009)
  [arXiv:0812.1328 [astro-ph]].

\bibitem{Bambi:2010hf}
  C.~Bambi and N.~Yoshida,
  %``Shape and position of the shadow in the $\delta = 2$ Tomimatsu-Sato space-time,''
  Class.\ Quant.\ Grav.\  {\bf 27}, 205006 (2010)
  [arXiv:1004.3149 [gr-qc]].
   
\bibitem{study1} A.~Abdujabbarov, F.~Atamurotov, Y.~Kucukakca, B.~Ahmedov and U.~Camci,
  %``Shadow of Kerr-Taub-NUT black hole,''
  Astrophys.\ Space Sci.\  {\bf 344} (2013) 429
%  doi:10.1007/s10509-012-1337-6
  [arXiv:1212.4949 [physics.gen-ph]].  
  
\bibitem{study2} F.~Atamurotov, A.~Abdujabbarov and B.~Ahmedov,
  %``Shadow of rotating Hořava-Lifshitz black hole,''
  Astrophys.\ Space Sci.\  {\bf 348} (2013) 179.
    
\bibitem{Moffat} J.~W.~Moffat,
  %``Modified Gravity Black Holes and their Observable Shadows,''
  Eur.\ Phys.\ J.\ C {\bf 75} (2015) no.3,  130
%  doi:10.1140/epjc/s10052-015-3352-6
  [arXiv:1502.01677 [gr-qc]].
  
\bibitem{quint2} A.~Abdujabbarov, B.~Toshmatov, Z.~Stuchlík and B.~Ahmedov,
  %``Shadow of the rotating black hole with quintessential energy in the presence of plasma,''
  Int.\ J.\ Mod.\ Phys.\ D {\bf 26} (2016) no.06,  1750051
%  doi:10.1142/S0218271817500511
  [arXiv:1512.05206 [gr-qc]].
  
\bibitem{study3} Z.~Younsi, A.~Zhidenko, L.~Rezzolla, R.~Konoplya and Y.~Mizuno,
  %``New method for shadow calculations: Application to parametrized axisymmetric black holes,''
  Phys.\ Rev.\ D {\bf 94} (2016) no.8,  084025
%  doi:10.1103/PhysRevD.94.084025
  [arXiv:1607.05767 [gr-qc]].

\bibitem{study4} P.~V.~P.~Cunha, C.~A.~R.~Herdeiro, B.~Kleihaus, J.~Kunz and E.~Radu,
  %``Shadows of Einstein–dilaton–Gauss–Bonnet black holes,''
  Phys.\ Lett.\ B {\bf 768} (2017) 373
%  doi:10.1016/j.physletb.2017.03.020
  [arXiv:1701.00079 [gr-qc]].

\bibitem{study5} M.~Wang, S.~Chen and J.~Jing,
  %``Shadow casted by a Konoplya-Zhidenko rotating non-Kerr black hole,''
  JCAP {\bf 1710} (2017) no.10,  051
%  doi:10.1088/1475-7516/2017/10/051
  [arXiv:1707.09451 [gr-qc]].

\bibitem{study6} H.~M.~Wang, Y.~M.~Xu and S.~W.~Wei,
  %``Shadows of Kerr-like black holes in a modified gravity theory,''
  JCAP {\bf 1903} (2019) no.03,  046
%  doi:10.1088/1475-7516/2019/03/046
  [arXiv:1810.12767 [gr-qc]].

\bibitem{bobir2017} B. Toshmatov, Z. Stuchlík, and B. Ahmedov Phys. Rev. D {\bf 95}, 084037 (2017)
\bibitem{Konoplya:2019sns}
  R.~A.~Konoplya,
  %``Shadow of a black hole surrounded by dark matter,''
  arXiv:1905.00064 [gr-qc]. Physics Letters B, in press (2019)
  \bibitem{sudipta2019}A. Mishra, S. Chakraborty, S. Sarkar, 	arXiv:1903.06376 [gr--qc]

 \bibitem{shakih2019b} R. Shakih,  arXiv:1904.08322 [gr-qc]

\bibitem{Panotopoulos:2017clp} 
  G.~Panotopoulos and \'A.~Rinc\'on,
  %``Stability of cosmic structures in scalar–tensor theories of gravity,''
  Eur.\ Phys.\ J.\ C {\bf 78}, no. 1, 40 (2018)

\bibitem{azreg2014} M. Azreg-A\"{i}nou, Phys. Rev. D {\bf 90}, 064041 (2014).

\bibitem{carter} B.~Carter,
  %``Global structure of the Kerr family of gravitational fields,''
  Phys.\ Rev.\  {\bf 174} (1968) 1559.

\bibitem{vazquez} S. Vazquez and E. Esteban, Nuovo Cim. {\bf 119}, 489 (2004).

\bibitem{ma} Zheng ZeMa,  Phys. Lett. B {\bf 666}, 376 (2008).

\bibitem{parikh1} M.K. Parikh, F. Wilczek, Phys. Rev. Lett. {\bf 85}, 5042 (2000).
 
\bibitem{parikh2} M.K. Parikh, Phys. Lett. B {\bf 546}, 189 (2002). 

\bibitem{parikh3} M.K. Parikh, Int. J. Mod. Phys. D {\bf 13}, 2351 (2004).

\bibitem{amir2016} A. Abdujabbarov, M. Amir, B. Ahmedov, and S.  Ghosh, Phys. Rev. D {\bf 93}, 104004 (2016).

\bibitem{papnoi2014} U. Papnoi, F.Atamurotov, S. Ghosh, and B. Ahmedov, Phys. Rev. D {\bf 90}, 024073 (2014).

\bibitem{misner}  C. W. Misner, K.S. Thorne, and J.A. Wheeler, Gravitation, (W.H. Freeman, San Francisco, 1973).

\bibitem{shao2013} Shao-Wen Wei and Yu-Xiao Liu JCAP {\bf 11}, 063 (2013).

\bibitem{hioki2009} K. Hioki and K.Maeda, Phys. Rev. D {\bf 80}, 024042 (2009).

\bibitem{spectrum} S.~W.~Hawking,
  %``Particle Creation by Black Holes,''
  Commun.\ Math.\ Phys.\  {\bf 43} (1975) 199
  Erratum: [Commun.\ Math.\ Phys.\  {\bf 46} (1976) 206].

\bibitem{wei2013} S. Wei and T. Liu, Journal of Cosmology and Astroparticles {\bf 11}, 063 (2013). 

\bibitem{Panotopoulos:2016wuu} G.~Panotopoulos and \'A.~Rinc\'on,
  %``Greybody factors for a nonminimally coupled scalar field in BTZ black hole background,''
  Phys.\ Lett.\ B {\bf 772}, 523 (2017).
  
\bibitem{Panotopoulos:2017yoe} G.~Panotopoulos and \'A.~Rinc\'on,
  %``Greybody factors for a minimally coupled massless scalar field in Einstein-Born-Infeld dilaton spacetime,''
  Phys.\ Rev.\ D {\bf 96}, no. 2, 025009 (2017).
  
\bibitem{Destounis:2018utr} K.~Destounis, G.~Panotopoulos and \'A.~Rinc\'on,
  %``Stability under scalar perturbations and quasinormal modes of 4D Einstein–Born–Infeld dilaton spacetime: exact spectrum,''
  Eur.\ Phys.\ J.\ C {\bf 78}, no. 2, 139 (2018).
  
\bibitem{Panotopoulos:2018pvu} G.~Panotopoulos and \'A.~Rinc\'on,
  %``Greybody factors for a minimally coupled scalar field in three-dimensional Einstein-power-Maxwell black hole background,''
  Phys.\ Rev.\ D {\bf 97}, no. 8, 085014 (2018).
  
\bibitem{Rincon:2018ktz} \'A.~Rinc\'on and G.~Panotopoulos,
  %``Greybody factors and quasinormal modes for a nonminimally coupled scalar field in a cloud of strings in (2+1)-dimensional background,''
  Eur.\ Phys.\ J.\ C {\bf 78}, no. 10, 858 (2018)      

\bibitem{lemos} J. Lemos, Class. Quantum Grav. {\bf 12}, 1081 (1995)

\bibitem{cai} R. Cai, Y. Zhang, Phys. Rev. D {\bf 54},4891 (1996).

\bibitem{Koch:2016uso} 
  B.~Koch, I.~A.~Reyes and \'A.~Rinc\'on,
  %``A scale dependent black hole in three-dimensional space–time,''
  Class.\ Quant.\ Grav.\  {\bf 33}, no. 22, 225010 (2016)
% doi:10.1088/0264-9381/33/22/225010
  [arXiv:1606.04123 [hep-th]].

% 2 %

\bibitem{Rincon:2017ypd} 
  \'A.~Rinc\'on, B.~Koch and I.~Reyes,
  %``BTZ black hole assuming running couplings,''
  J.\ Phys.\ Conf.\ Ser.\  {\bf 831}, no. 1, 012007 (2017)
% doi:10.1088/1742-6596/831/1/012007
  [arXiv:1701.04531 [hep-th]].

% 3 %

\bibitem{Rincon:2017goj} 
  \'A.~Rinc\'on, E.~Contreras, P.~Bargue\~no, B.~Koch, G.~Panotopoulos and A.~Hern\'andez-Arboleda,
  %``Scale dependent three-dimensional charged black holes in linear and non-linear electrodynamics,''
  Eur.\ Phys.\ J.\ C {\bf 77}, no. 7, 494 (2017)
% doi:10.1140/epjc/s10052-017-5045-9
  [arXiv:1704.04845 [hep-th]].

% 4 %

\bibitem{Rincon:2017ayr} 
  \'A.~Rinc\'on and B.~Koch,
  %``On the null energy condition in scale dependent frameworks with spherical symmetry,''
  J.\ Phys.\ Conf.\ Ser.\  {\bf 1043}, no. 1, 012015 (2018)
% doi:10.1088/1742-6596/1043/1/012015
  [arXiv:1705.02729 [hep-th]].

% 5 %

\bibitem{Contreras:2017eza} 
  E.~Contreras, \'A.~Rinc\'on, B.~Koch and P.~Bargue\~no,
  %``A regular scale-dependent black hole solution,''
  Int.\ J.\ Mod.\ Phys.\ D {\bf 27}, no. 03, 1850032 (2017)
% doi:10.1142/S0218271818500323
  [arXiv:1711.08400 [gr-qc]].

% 6 %

\bibitem{Rincon:2018sgd} 
  \'A.~Rinc\'on and G.~Panotopoulos,
  %``Quasinormal modes of scale dependent black holes in (1 + 2)-dimensional Einstein-power-Maxwell theory,''
  Phys.\ Rev.\ D {\bf 97}, no. 2, 024027 (2018)
% doi:10.1103/PhysRevD.97.024027
  [arXiv:1801.03248 [hep-th]].

% 7 %

\bibitem{Hernandez-Arboleda:2018qdo} 
  A.~Hern\'andez-Arboleda, \'A.~Rinc\'on, B.~Koch, E.~Contreras and P.~Bargue\~no,
  %``Preliminary test of cosmological models in the scale-dependent scenario,''
  arXiv:1802.05288 [gr-qc].

% 8 %

\bibitem{Contreras:2018dhs} 
  E.~Contreras, \'A.~Rinc\'on, B.~Koch and P.~Bargue\~no,
  %``Scale-dependent polytropic black hole,''
  Eur.\ Phys.\ J.\ C {\bf 78}, no. 3, 246 (2018)
% doi:10.1140/epjc/s10052-018-5709-0
  [arXiv:1803.03255 [gr-qc]].

% 9 %

\bibitem{Rincon:2018dsq} 
  \'A.~Rinc\'on, E.~Contreras, P.~Bargue\~no, B.~Koch and G.~Panotopoulos,
  %``Scale-dependent ( $2+1$ )-dimensional electrically charged black holes in Einstein-power-Maxwell theory,''
  Eur.\ Phys.\ J.\ C {\bf 78}, no. 8, 641 (2018)
% doi:10.1140/epjc/s10052-018-6106-4
  [arXiv:1807.08047 [hep-th]].

% 10 %

\bibitem{Contreras:2018gct} 
  E.~Contreras, \'A.~Rinc\'on and J.~M.~Ram\'irez-Velasquez,
  %``Relativistic dust accretion onto a scale--dependent polytropic black hole,''
  Eur.\ Phys.\ J.\ C {\bf 79}, no. 1, 53 (2019)
% doi:10.1140/epjc/s10052-019-6601-2
  [arXiv:1810.07356 [gr-qc]].

% 11 %

\bibitem{Canales:2018tbn} 
  F.~Canales, B.~Koch, C.~Laporte and \'A.~Rinc\'on,
  %``Vacuum energy density: deflation during inflation,''
  arXiv:1812.10526 [gr-qc].

% 12 %

\bibitem{Rincon:2019cix} 
  \'A.~Rinc\'on, E.~Contreras, P.~Bargue\~no and B.~Koch,
  %``Scale-dependent planar Anti-de Sitter black hole,''
  arXiv:1901.03650 [gr-qc].

% 13 %

\bibitem{Rincon:2019zxk} 
  \'A.~Rinc\'on and J.~R.~Villanueva,
  %``The Sagnac effect on a scale-dependent rotating BTZ black hole background,''
  arXiv:1902.03704 [gr-qc].

% 14 %

\bibitem{Contreras:2019fwu} 
  E.~Contreras, \'A.~Rinc\'on and P.~Bargue\~no,
  %``Five-Dimensional Scale-Dependent Black Holes with Constant Curvature and Solv Horizons,''
  arXiv:1902.05941 [gr-qc].

\bibitem{Contreras:2018gpl} 
  E.~Contreras and P.~Bargue\~no,
  %``Scale--dependent Hayward black hole and the generalized uncertainty principle,''
  Mod.\ Phys.\ Lett.\ A {\bf 33}, no. 32, 1850184 (2018)


\bibitem{Contreras:2018swc} 
  E.~Contreras and P.~Bargue\~no,
  %``A self-sustained traversable scale-dependent wormhole,''
  Int.\ J.\ Mod.\ Phys.\ D {\bf 27}, no. 09, 1850101 (2018)


\end{thebibliography}
\end{document}